\documentclass[aps,showpacs,nofootinbib,pre,superscriptaddress,floatfix,12pt]{revtex4-1}
     
\usepackage{multirow}
\usepackage{amsfonts}
\usepackage{amsmath}
\usepackage{amssymb}
\usepackage{graphicx}
\usepackage{dcolumn}
\usepackage{times}
\usepackage{color}
\usepackage[a4paper]{geometry}
\usepackage{hhline}

\begin{document}

\title{Super-Lagrangian and  variational principle for generalized continuity equations}

\author{F.~K.~Diakonos}
\email[]{fdiakono@phys.uoa.gr}
\affiliation{Department of Physics, University of Athens, GR-15771 Athens, Greece}

\author{P.~Schmelcher}
\email[]{pschmelc@physnet.uni-hamburg.de}
\affiliation{Zentrum f\"{u}r Optische Quantentechnologien, Universit\"{a}t Hamburg, Luruper Chaussee 149, 22761 Hamburg, Germany}
\affiliation{The Hamburg Centre for Ultrafast Imaging, Universit\"{a}t Hamburg, Luruper Chaussee 149, 22761 Hamburg, Germany}
\date{\today}

\begin{abstract} 
We present a variational approach which shows that the wave functions belonging to quantum systems in different potential landscapes, are pairwise linked to each other through a generalized continuity equation. This equation contains a source term proportional to the potential difference. In case the potential landscapes are related by a linear symmetry transformation in a finite domain of the embedding space, the derived continuity equation leads to generalized currents which are divergence free within this spatial domain. In a single spatial dimension these generalized currents are invariant. In contrast to the standard continuity equation, originating from the abelian $U(1)$-phase symmetry of the standard Lagrangian, the generalized continuity equations derived here, are based on a non-abelian $SU(2)$-transformation of a Super-Lagrangian. Our approach not only provides a rigorous theoretical framework to study quantum mechanical systems in potential landscapes possessing local symmetries, but it also reveals a general duality between quantum states corresponding to different Schr\"{o}dinger problems.

\end{abstract}
\maketitle

\section{Introduction}\label{sec:1}
The continuity equation obeyed by the probability density and currents of a closed quantum system lies at the heart of the quantum description of
nature \cite{Messiah2014}. It is a consequence of probability conservation and can be derived from basic principles, applying Noethers theorem \cite{Noether1918,Neuenschwander2011} to a suitably chosen Lagrangian. In fact, the continuity equation is directly related to an internal symmetry of Hilbert space, expressed by the
freedom of the choice of the phase of quantum states associated with unitarity \cite{Messiah2014,Schwarzbach2011}. An interesting question in this context is whether, specific spatio-temporal coordinate transformations, leaving the Hamiltonian of a quantum system invariant, may have a distinct impact on the corresponding continuity equations. An important step towards answering this question has been performed in a recent work \cite{Spourdalakis2016} where it was shown that in the aforementioned case, symmetry-induced bilocal continuity equations for Hermitian and non-Hermitian quantum mechanical systems exist. An interesting feature of these continuity equations is that they are special cases of generalized continuity equations involving two distinct wave-fields, derived from a corresponding two-field Lagrangian which is invariant under both phase and dilatation transformations of these fields. As mentioned in \cite{Spourdalakis2016}, the currents in the derived continuity equations are interpreted as correlators between the two involved wave fields, thereby generalizing the concept of the standard probability currents. Thus, the generalized continuity equations derived in \cite{Spourdalakis2016} provide the appropriate theoretical framework to describe the impact of global discrete symmetries on properties related to the unitarity of the associated Schr\"{o}dinger states. In contrast to the above-addressed case of global symmetries, we meet in physical systems often the situation where symmetries are not globally valid but they hold only in finite spatial domains, which we term as local symmetries. Recently it has been shown \cite{Kalozoumis2014} that a new class of non-local currents occurs in systems with local symmetries. They are induced by the local symmetries of the potential term in the Schr\"{o}dinger equation, and remain spatially invariant within the finite domains of local symmetry. The generalized continuity equations derived in \cite{Spourdalakis2016} focus on the case of global symmetry and do not apply to this situation. In the present work we fill this gap and provide a theoretical framework which allows to derive the generalized continuity equations obeyed by the local symmetry induced non-local currents. It is based on a variational approach utilizing a Super-Lagrangian.

In fact, in the present work we generalize the variational approach developed in ref. \cite{Spourdalakis2016} in a twofold way: (i) we introduce a Super-Lagrangian which is constructed from the individual Lagrangians describing the dynamics of a quantum system in two different potential landscapes and (ii) based on an $SU(2)$-transformation of the wave-fields in the Super-Lagrangian we derive a generalized continuity equation containing a source term proportional to the difference of the two potentials. This continuity equation involves a novel class of correlator currents which combine eigenstates i.e. solutions of the two different individual Schr\"{o}dinger problems, resulting variationally from the Super-Lagrangian. We show that these correlator-currents, together with their generalized continuity equations
provide a distinct platform to variationally derive conservation laws, that originate from symmetries of the interaction potential valid in restricted spatial domains. The latter are the aforementioned local symmetries which have been in the focus of a number of recent investigations \cite{LocSym}. Furthermore, we show that combining $N$-Schr\"{o}dinger problems in a single Super-Lagrangian, which in turn allows the application of an $SU(N)$ internal symmetry transformation to the wave-field multiplet, leads to generalized continuity equations of a correspondingly generalized form. In fact, we obtain a set of $\frac{N(N-1)}{2}$ continuity equations, similar to those of the $SU(2)$-case, each one containing pairwise combinations of 
wave-fields corresponding to solutions of the individual Schr\"{o}dinger problems. Finally, we make the link between the correlator currents originating from the Super-Lagrangian to the non-local invariant currents obtained in ref. \cite{Kalozoumis2014}, in the context of the quantum states in potentials with local symmetries.

Our work is organized as follows. In section II we introduce the aforementioned Super-Lagrangian consisting of two individual Schr\"{o}dinger-type Lagrangians and we derive the associated generalized continuity equation with a source term. In Section III we extend the Super-Lagrangian such that it contains $N$ individual Schr\"{o}dinger-type Lagrangians and we demonstrate that an $SU(N)$-transformation of the involved wave-fields leads to ${N(N-1) \over 2}$ different generalized continuity equations of the $SU(2)$ form. In section IV we show that for a particular choice of the potentials and their symmetries in the two Lagrangians, constituting the Super-Lagrangian in the $SU(2)$ case, we obtain the spatially constant non-local currents derived in ref. \cite{Kalozoumis2014} for the case of locally symmetric potentials. We provide also an example concerning the application of our approach to a concrete quantum system evolving in a potential with specific symmetry properties. Finally, in section V we present our concluding remarks containing a summary of our results as well as an outlook.

\section{Super-Lagrangian and generalized continuity equations}

For simplicity we begin our analysis by considering the dynamics of a single quantum particle in one spatial dimension. Let us assume that the quantum particle evolves in an external potential $V_1(x)$. Then, it is straightforward to construct a Lagrangian density $\mathcal{L}_1$ which leads variationally to the corresponding Schr\"{o}dinger equation. A possible choice is 
\begin{eqnarray}
\mathcal{L}_1 = \mathcal{L}_1 \left[ \Psi_1 \right]= \Psi_1(x,t)^{\dagger} [i \partial_t - \hat{H}_0 - V_1(x)] \Psi_1(x,t) + \mathrm{h.c.}\\ \nonumber
\Rightarrow [i \partial_t - \hat{H}_0 - V_1(x)] \Psi_1(x,t)=0
\label{eq:1}
\end{eqnarray}
with  $\hat{H}_0=-\frac{1}{2}\frac{\partial^2}{\partial x^2}$ and $\Psi_1(x,t)$ is the wave-field which yields a solution of the corresponding Schr\"{o}dinger equation. Notice that we set $\hbar=1$, $m=1$ throughout this work. If the external potential is a different one, e.g. $V_2(x)$, then we have to replace in all expressions occurring in (\ref{eq:1}) the index $1$ by the index $2$. Thus, for a given quantum system with specifying its potential term there is a one-to-one correspondence to the associated Lagrangian. An alternative way to obtain variationally the two Schr\"{o}dinger equations corresponding to the two Schr\"{o}dinger-type problems with potential energy terms $V_1(x)$ and $V_2(x)$ respectively, is to define first a single Lagrangian $\mathcal{L}$, being the sum of the two individual Lagrangians $\mathcal{L}_1$ and $\mathcal{L}_2$, as
\begin{equation}
\mathcal{L} \left[ \Psi_1, \Psi_2 \right] = \mathcal{L}_1 \left[ \Psi_1 \right]  + \mathcal{L}_2  \left[ \Psi_2 \right]
\label{eq:2}
\end{equation}
and apply subsequently the standard variational approach based on variations of the two wave-fields $\Psi_1(x,t)$ and $\Psi_2(x,t)$ involved in $\mathcal{L}$. This appears at a first look as a trivial generalization. However there are some subtleties connected to this step which give rise to new properties as we will show in the following. The first subtle point is the fact that the Lagrangian (\ref{eq:2}) refers to a single quantum particle or quantum system with two possible dynamical evolutions, determined by the different potential landscapes $V_1(x)$ and $V_2(x)$, and not to two different quantum subsystems of a larger system. In the latter case the wave-field of the entire system would depend on two variables $x_1$ and $x_2$ characterizing each subsystem separately. To explore some further subtle properties contained in the formulation (\ref{eq:2}) we rewrite $\mathcal{L}$ in a matrix form    
\begin{equation}
\mathcal{L}=\begin{pmatrix}\Psi_1(x,t) \\ \Psi_2(x,t) \end{pmatrix}^{\dagger} \begin{bmatrix} i \partial_t - \hat{H}_0 - V_1(x) & 0 \\ 0 & i \partial_t - \hat{H}_0 - V_2(x) \end{bmatrix}  \begin{pmatrix}\Psi_1(x,t) \\ \Psi_2(x,t) \end{pmatrix} + \mathrm{h.c.}
\label{eq:3}
\end{equation}
which can be decomposed as
\begin{equation}
\mathcal{L} \left[ \Psi_1, \Psi_2 \right] = \mathcal{L}_0 \left[ \Psi_1, \Psi_2 \right] + \mathcal{L}_I \left[ \Psi_1, \Psi_2 \right]
\label{eq:4}
\end{equation}
using the notation
\begin{eqnarray}
\mathcal{L}_0 &=& \begin{pmatrix}\Psi_1(x,t) \\ \Psi_2(x,t) \end{pmatrix}^{\dagger} \begin{bmatrix} i \partial_t - \hat{H}_0  & 0 \\ 0 & i \partial_t - \hat{H}_0  \end{bmatrix}  \begin{pmatrix}\Psi_1(x,t) \\ \Psi_2(x,t) \end{pmatrix}  + \mathrm{h.c.} \nonumber \\
\mathcal{L}_I &=& 
\begin{pmatrix}\Psi_1(x,t) \\ \Psi_2(x,t) \end{pmatrix}^{\dagger} \begin{bmatrix} - V_1(x)  & 0 \\ 0 &  -V_2(x)  \end{bmatrix}  \begin{pmatrix}\Psi_1(x,t) \\ \Psi_2(x,t) \end{pmatrix}  + \mathrm{h.c.}
\label{eq:5}
\end{eqnarray}
The second term $\mathcal{L}_I$ in Eq.~(\ref{eq:5}) can be written
as
\begin{equation}
\mathcal{L}_I = - \begin{pmatrix}\Psi_1(x,t) \\ \Psi_2(x,t) \end{pmatrix}^{\dagger} \left[ V_1(x)
 \frac{\hat{\mathbf{I}} + \hat{\sigma}_z}{2} + V_2(x) \frac{\hat{\mathbf{I}} - \hat{\sigma}_z}{2} \right] \begin{pmatrix}\Psi_1(x,t) \\ \Psi_2(x,t) \end{pmatrix} + \mathrm{h.c.}
\label{eq:6}
\end{equation}
with $\hat{\sigma}_z$ being the corresponding Pauli matrix. In fact the term proportional to the identity operator $\hat{\mathbf{I}}$ can be absorbed in $\mathcal{L}_0$ allowing us to rewrite the Lagrangian (\ref{eq:4}) as follows
\begin{equation}
\mathcal{L}=\mathcal{L}_{id} + \mathcal{L}_{dif}
\label{eq:7}
\end{equation}
with
\begin{eqnarray}
\mathcal{L}_{id} &=& \begin{pmatrix}\Psi_1(x,t) \\ \Psi_2(x,t) \end{pmatrix}^{\dagger} \left[i \partial_t - \hat{H}_0 - \frac{V_1(x)+V_2(x)}{2} \right] \hat{\mathbf{I}} \begin{pmatrix}\Psi_1(x,t) \\ \Psi_2(x,t) \end{pmatrix} + \mathrm{h.c.} \\ \label{eq:8}
\mathcal{L}_{dif} &=& \begin{pmatrix}\Psi_1(x,t) \\ \Psi_2(x,t) \end{pmatrix}^{\dagger} \left( \frac{V_2(x) - V_1(x)}{2} \right) \hat{\sigma}_z \begin{pmatrix}\Psi_1(x,t) \\ \Psi_2(x,t) \end{pmatrix}  + \mathrm{h.c.} \label{eq:9}
\end{eqnarray}
Clearly the term $\mathcal{L}_{id}$ in the Lagrangian (\ref{eq:7}) is invariant with respect to $SU(2)$ transformations of the type
\begin{equation}
\begin{pmatrix} \Psi_1(x,t) \\ \Psi_2(x,t) \end{pmatrix}
\Longrightarrow \exp[i \mathbf{a} \cdot \frac{\hat{\mathbf{\sigma}}}{2}] \begin{pmatrix} \Psi_1(x,t) \\ \Psi_2(x,t) \end{pmatrix}
\label{eq:10}
\end{equation}
while the term (\ref{eq:9}) is not. Furthermore, for two potentials which obey $V_2(x)=V_1(F(x))$ with a linear coordinate
transformation $x \longrightarrow F(x)= \sigma x + \rho; \sigma = \pm 1, \rho \in \mathbb{R}$ (parity, translation) the resulting 
Schr{\"o}dinger equation is invariant under this transformation since $\hat{H}_0$ is invariant too.

We will in the following explore the variation of $\mathcal{L}$ under an infinitesimal $SU(2)$ transformation of the structure given in
(\ref{eq:10}). If $\mathcal{L}$ was invariant with respect to the transformation (\ref{eq:10}), according to Noether's theorem, we would obtain a continuity equation associated with this invariance. Here our aim is to determine, whether the absence of the invariance of $\mathcal{L}$ still allows to derive correspondingly generalized continuity equations. In our treatment we will use $\mathbf{a}=(\delta \theta_1,\delta \theta_2,0)$ without loss of generality, since the variation proportional to the $\hat{\sigma}_z$-term in the infinitesimal transformation, vanishes. As already mentioned for
\begin{equation}
\begin{pmatrix} \Psi_1(x,t) \\ \Psi_2(x,t) \end{pmatrix}
\Longrightarrow \left[ \hat{\mathbf{I}} + \frac{i}{2} (\delta\theta_1 \hat{\sigma}_x + \delta \theta_2 \hat{\sigma}_y) \right] \begin{pmatrix} \Psi_1(x,t) \\ \Psi_2(x,t) \end{pmatrix}
\label{eq:11}
\end{equation}
the term $\mathcal{L}_{id}$ remains invariant, i.e. $\delta \mathcal{L}_{id}=0$ and we have to consider only the variation of the term
$\mathcal{L}_{dif}$. The latter becomes
\begin{eqnarray}
\delta \mathcal{L}_{dif}=(V_2(x)-V_1(x))& & \left[ i (\Psi_2^*(x,t) \Psi_1(x,t) - \Psi_2(x,t) \Psi_1^*(x,t)) \delta \theta_1 \right. \nonumber \\
& &- \left. (\Psi_2^*(x,t) \Psi_1(x,t) + \Psi_2(x,t) \Psi_1^*(x,t)) \delta \theta_2 \right]
\label{eq:12}
\end{eqnarray}
Thus we have 
\begin{equation}
\delta \mathcal{L}=\delta \mathcal{L}_{dif}
\label{eq:13}
\end{equation}
On the other hand the variation of $\mathcal{L}$ can be written as
\begin{equation}
\delta \mathcal{L} = \sum_{i=1}^2 \left( \frac{\partial \mathcal{L}}{\partial \Psi_i} \delta \Psi_i + \frac{\partial \mathcal{L}}{\partial(\partial_t \Psi_i)} \delta (\partial_t \Psi_i) + \frac{\partial \mathcal{L}}{\partial(\partial_x \Psi_i)} \delta (\partial_x \Psi_i) + \mathrm{h.c.} \right)
\label{eq:14}
\end{equation} 
Using 
\begin{equation}
\delta (\partial_t \Psi_i)= \partial_t (\delta \Psi_i)~~~;~~~
\delta (\partial_x \Psi_i)= \partial_x (\delta \Psi_i)
\label{eq:15}
\end{equation}
and similarly for $\Psi_i^*$, we obtain
\begin{eqnarray}
\delta \mathcal{L} &=& \sum_{i=1}^2 \left( \left[\frac{\partial \mathcal{L}}{\partial \Psi_i} - \partial_t \left(\frac{\partial \mathcal{L}}{\partial(\partial_t \Psi_i)}\right) - \partial_x \left(\frac{\partial \mathcal{L}}{\partial(\partial_x \Psi_i)}\right) \right] \delta \Psi_i \right. \nonumber \\
 &+& \left.  \partial_t \left(\frac{\partial \mathcal{L}}{\partial(\partial_t \Psi_i)} \delta \Psi_i \right) + \partial_x \left(\frac{\partial \mathcal{L}}{\partial(\partial_x \Psi_i)} \delta \Psi_i \right) + \mathrm{h.c.} \right)
\label{eq:16}
\end{eqnarray}
where the first term in the brackets vanishes if we use the equations of motion for the wave fields. The second and third terms form the divergence of a four-current as it appears in the usual continuity equation. Let us derive it explicitly.
The variations of the fields are
\begin{eqnarray}
\delta \Psi_1(x,t) &=& \frac{i}{2} (\delta \theta_1 - i \delta \theta_2) \Psi_2(x,t)~~~~;~~~~\delta \Psi_1^*(x,t) = -\frac{i}{2} (\delta \theta_1 + i \delta \theta_2) \Psi_2^*(x,t) 
\nonumber \\
\delta \Psi_2(x,t) &=& \frac{i}{2} (\delta \theta_1 + i \delta \theta_2) \Psi_1(x,t)~~~~;~~~~\delta \Psi_2^*(x,t) = -\frac{i}{2} (\delta \theta_1 - i \delta \theta_2) \Psi_1^*(x,t)
\label{eq:17}
\end{eqnarray}
while the derivatives of the Lagrangian give
\begin{eqnarray}
\frac{\partial \mathcal{L}}{\partial(\partial_t \Psi_i)} &=&i\Psi_i^*~~~~~~~~~~;~~~~~~\frac{\partial \mathcal{L}}{\partial(\partial_t \Psi_i^*)} = -i \Psi_i \nonumber \\
\frac{\partial \mathcal{L}}{\partial(\partial_x \Psi_i)} &=&- \partial_x \Psi_i^*~~~~~;~~~~~~\frac{\partial \mathcal{L}}{\partial(\partial_x \Psi_i^*)} = - \partial_x \Psi_i
\label{eq:18}
\end{eqnarray}
It must be noticed that the derivation of the derivatives in the second line of Eq.~(\ref{eq:18}) requires a partial integration in the action, which can be avoided if we write $\hat{H}_0= \frac{1}{2} \frac{\overleftarrow{\partial}}{\partial x} \frac{\overrightarrow{\partial}}{\partial x}$ instead of $\hat{H}_0=-\frac{1}{2}\frac{\partial^2}{\partial x^2}$ (where the arrows determine in which direction the corresponding
derivative operator acts). Inserting Eqs.~(\ref{eq:17}), (\ref{eq:18}) into the second and third term of Eq.~(\ref{eq:16}) and using the expression (\ref{eq:14}) we determine $\delta \mathcal{L}$. Then, employing  Eq.~(\ref{eq:13}) and replacing $\delta \mathcal{L}_{dif}$ by the rhs of Eq.~(\ref{eq:12}) we obtain:
\begin{eqnarray}
\partial_t \left( \mathrm{Re}\Psi_1 \Psi_2^* \right) - \frac{1}{2} \partial_x \left[\mathrm{Im} \left( \Psi_1 \partial_x \Psi_2^* - \Psi_2^* \partial_x \Psi_1 \right) \right] &=& (V_2(x)-V_1(x))\mathrm{Im} \Psi_1 \Psi_2^* \nonumber \\
\partial_t \left( \mathrm{Im}\Psi_1 \Psi_2^* \right) + \frac{1}{2} \partial_x \left[\mathrm{Re} \left( \Psi_1 \partial_x \Psi_2^* - \Psi_2^* \partial_x \Psi_1 \right) \right] &=& -(V_2(x)-V_1(x))\mathrm{Re} \Psi_1 \Psi_2^* 
\label{eq:19}
\end{eqnarray}
The first of these equations results from the terms of order $\delta \theta_1$ while the second stems from terms of order $\delta \theta_2$. Eqs.~(\ref{eq:19}) can be combined to a generalized continuity equation with a source term 
\begin{equation}
\partial_t \left(\Psi_1 \Psi_2^* \right) + \frac{1}{2i} \partial_x  \left( \Psi_2^* \partial_x \Psi_1 - \Psi_1 \partial_x \Psi_2^* \right) = i (V_2(x)-V_1(x))\Psi_1 \Psi_2^* 
\label{eq:20}
\end{equation}
which contains wave-fields obeying different Schr\"{o}dinger equations. The discussion so far has focused on a single particle in an external potential. It is straightforward to extend the obtained result to higher dimensions repeating the procedure described previously. The equations in higher dimensions are obtained by replacing $x$ with $\mathbf{x}$ and $\partial_x$ with $\mathbf{\nabla}$ in Eq.~(\ref{eq:20}). For a system of $N$ identical interacting particles exposed to an external potential in $d$ spatial dimensions, and following the above derivations described for a single particle in one spatial dimension, one obtains a continuity equation which is the same as in Eq.~(\ref{eq:20}) up to the replacement of $x$ by $\mathbf{R}=(x_{1,1},x_{1,2},..., x_{1,d}, x_{2,1}, x_{2,2},..., x_{2,d},..., x_{N,1}, x_{N,2},..., x_{N,d})$ and $\partial_x$ by $\mathbf{\nabla_R}$. The super-vector $\mathbf{R}$ is constructed from all the $d$-dimensional position vectors of each particle, using the notation $x_{i,j}$ for its entries, where $i$ ($i=1, 2, .., N$) is the particle index and $j$ the index indicating the corresponding spatial components ($j=1, 2,..., d$). In an analogous way we construct the super-gradient $\mathbf{\nabla_R}$. Thus, Eq.~(\ref{eq:20}) generalizes to the continuity equation:
\begin{equation}
\partial_t \left(\Psi_1 \Psi_2^* \right) + \frac{1}{2i} \mathbf{\nabla_R} \left( \Psi_2^* \mathbf{\nabla_R} \Psi_1 - \Psi_1 \mathbf{\nabla_R} \Psi_2^* \right) = i (V_2(\mathbf{R})-V_1(\mathbf{R}))\Psi_1 \Psi_2^* 
\label{eq:21}
\end{equation}
containing a source term which relates the wave fields of two arbitrary Schr\"{o}dinger problems containing the same degrees of freedom.

\section{$N$-state Super-Lagrangian for $SU(N)$ symmetry}

In this section we explore the extension of the Super-Lagrangian, introduced in the previous section, to the case when the involved wave fields correspond to $N$ Schr\"{o}dinger problems concerning a quantum system in $N$ different potential landscapes. Following the treatment in section II we write
\begin{eqnarray}
\mathcal{L}^{(N)}&=&\begin{pmatrix}\Psi_1(x,t) \\ \Psi_2(x,t) \\ . \\ . \\ . \\ \Psi_N(x,t) \end{pmatrix}^{\dagger} \begin{bmatrix} i \partial_t - \hat{H}_0 - V_1(x) & 0 & . & . & . & 0\\ 0 & i \partial_t - \hat{H}_0 - V_2(x) & 0 & . & . &0 \\ . & . & . & . & . & . \\  . & . & . & . & . & . \\  . & . & . & . & . & . \\ 0 & . & . & . & 0 & i \partial_t - \hat{H}_0 - V_N(x) \end{bmatrix}  \begin{pmatrix}\Psi_1(x,t) \\ \Psi_2(x,t) \\ . \\ . \\ . \\ \Psi_N(x,t) \end{pmatrix} \nonumber \\
&+& \mathrm{h.c.}
\label{eq:22}
\end{eqnarray}
For convenience we will use in the following the notation
\begin{equation}
\mathbf{\Psi}^{(N)}(x,t)=\begin{pmatrix}\Psi_1(x,t) \\ \Psi_2(x,t) \\ . \\ . \\ . \\ \Psi_N(x,t) \end{pmatrix}
\label{eq:23}
\end{equation} 
Our goal is to extend the derivation of the generalized continuity equation, performed in the previous section for $N=2$, to this general $N$-dimensional case. The Lagrangian in Eq.~(\ref{eq:23}) can be written in the form dictated by Eq.~(\ref{eq:7})
\begin{eqnarray}
\mathcal{L}^{(N)}&=&\mathcal{L}^{(N)}_{id} + \mathcal{L}^{(N)}_{dif} \nonumber \\
\mathcal{L}^{(N)}_{id} &=& \mathbf{\Psi}^{(N)}(x,t)^{\dagger} \left[i \partial_t - \hat{H}_0 - \frac{1}{N} \sum_{i=1}^N V_i(x) \right] \hat{\mathbf{I}} \mathbf{\Psi}^{(N)}(x,t) + \mathrm{h.c.} \nonumber \\
\mathcal{L}^{(N)}_{dif} &=& \mathbf{\Psi}^{(N)}(x,t)^{\dagger} \sum_{k=1}^{N-1} c_k(x) \hat{\mathbf{D}}_k \mathbf{\Psi}^{(N)}(x,t) + \mathrm{h.c.}
\label{eq:24}
\end{eqnarray}
where $\hat{\mathbf{D}}_k$ ($k=1, 2,..., N-1$) are the $SU(N)$-generators which form the corresponding Cartan subalgebra given as
\begin{equation}
\hat{\mathbf{D}}_k=\sqrt{\frac{2}{k(k+1)}} \begin{bmatrix} 1 & 0 & 0 & 0 & ... & 0 & 0 \\ 0 & 1 & 0 & 0 & ... & 0 & 0 \\ ... & ... & ... & ... & ... & ... & ... \\  0 & 0 & ... & 1 & 0 & 0 & ... \\  0 & 0& ... & 0 & -k & 0 & ... \\ ... & ... & ... & ... & ... & 0 & ... \\ 0 & 0 & ... & 0 & 0 & ... & 0 \end{bmatrix}_{N \times N}~~;~~~~~k=1, 2, ..., N-1
\label{eq:25}
\end{equation}

It is straightforward to derive the coefficients $c_k(x)$ yielding
\begin{equation}
c_k(x)=\sqrt{\frac{k+1}{2 k}} \left[ V_{k+1}(x) - \displaystyle{\sum_{i=1}^{k+1}} V_i(x) \right]~~~;~~~k=1, 2,..., N-1
\label{eq:26}
\end{equation}  
Reminiscent of the case $N=2$, the first part of the Super-Lagrangian $\mathcal{L}^{(N)}_{id}$ is invariant under $SU(N)$-transformations of the wave field while the second part $\mathcal{L}^{(N)}_{dif}$ is not. Therefore, we perform $SU(N)$ transformations of $\mathbf{\Psi}^{(N)}(x,t)$ and derive the corresponding variation $\delta \mathcal{L}^{(N)}_{dif}$ in an analogy to the treatment in the previous section. Obviously, we do need to consider only transformations of the type
\begin{equation}
\mathbf{\Psi}^{(N)}(x,t) \Longrightarrow \exp[i \mathbf{a} \cdot \frac{\hat{\mathbf{\Lambda}}}{2} ] \mathbf{\Psi}^{(N)}(x,t)
\label{eq:27}
\end{equation}
where the vector $\hat{\mathbf{\Lambda}}$ consists of generators of $SU(N)$ which do not belong to the corresponding Cartan subalgebra. To represent these generators, we make use of the operators $\hat{E}^{[i,j]}$ with $i < j$, defined as $N \times N$ matrices with zero entries, except of the element $\hat{E}^{[i,j]}_{i,j}$ which is equal to $1$. Clearly, $\hat{E}^{[i,j]}$ are not hermitian and therefore they do not represent $SU(N)$ generators (they do not obey the associated Lie algebra), nevertheless, they provide a suitable basis for expressing these generators. From each $\hat{E}^{[i,j]}$ one can construct two hermitian (traceless) matrices $\hat{\lambda}_{1}^{[i,j]}$ and $\hat{\lambda}_2^{[i,j]}$ as follows
\begin{equation}
\hat{\lambda}_1^{[i,j]}=\hat{E}^{[i,j]}  + (\hat{E}^{[i,j]})^{\dagger}~~~~~~;~~~~~~\hat{\lambda}_2^{[i,j]}=i (\hat{E}^{[i,j]}  - (\hat{E}^{[i,j]})^{\dagger})
\label{eq:28}
\end{equation}
Since there are $N(N-1)/2$ independent matrices $\hat{E}^{[i,j]}$ the set of matrices $\hat{\lambda}^{[i,j]}_k$ with $k=1, 2$ has cardinality $N(N-1)$ and represents all generators of $SU(N)$ which do not belong to the corresponding Cartan subalgebra. Thus, the most general infinitesimal $SU(N)$ transformation we need to consider for our purpose, has the form
\begin{equation}
\mathbf{\Psi}^{(N)}(x,t) \Longrightarrow \left[ \hat{I} + i \displaystyle{\sum_{\stackrel{i,j}{i < j}}} \left( \delta \theta_{i,j} \hat{\lambda}_1^{[i,j]} + \delta \phi_{i,j} \hat{\lambda}_2^{[i,j]}\right)\right] \mathbf{\Psi}^{(N)}(x,t)
\label{eq:29}
\end{equation} 
In fact our analysis, as one can easily show, can be restricted to the transformation
\begin{equation}
\mathbf{\Psi}^{(N)}(x,t) \Longrightarrow \left[ \hat{I} + i \left( \delta \theta \hat{\lambda}_1^{[i,j]} + \delta \phi \hat{\lambda}_2^{[i,j]}\right)\right] \mathbf{\Psi}^{(N)}(x,t)
\label{eq:30}
\end{equation} 
without loss of generality, since they represent the complete set of$~SU(N)$ generators which do not commute with a given $\hat{D}$. Introducing the operator $\hat{O}(x)=\sum_{i=1}^{N-1} c_i(x) \hat{D}_i$ it is straightforward to show the relations
\begin{equation}
[\hat{O}(x), \hat{\lambda}^{[i,j]}_1]= i (V_i(x)-V_j(x)) \hat{\lambda}^{[i,j]}_2~~~~~;~~~~~[\hat{O}(x), \hat{\lambda}^{[i,j]}_2]= -i (V_i(x)-V_j(x)) \hat{\lambda}^{[i,j]}_1
\label{eq:31}
\end{equation}
Based on Eqs.~(\ref{eq:31}) one can repeat the procedure followed in the previous section to obtain a generalized continuity equation of the form
\begin{equation}
\partial_t \left(\Psi_i \Psi_j^* \right) + \frac{1}{2i} \partial_x  \left( \Psi_j^* \partial_x \Psi_i - \Psi_i \partial_x \Psi_j^* \right) = i (V_j(x)-V_i(x))\Psi_i \Psi_j^* 
\label{eq:32}
\end{equation}
In fact there are $N(N-1)/2$ different choices for the matrices $\hat{\lambda}^{[i,j]}$ which correspond to the $N(N-1)/2$ different pairs $(V_j(x)-V_i(x))$ occurring as coefficients on the right hand side of the generalized continuity equation. Thus we end up with $N(N-1)/2$ different generalized continuity equations of the type given in Eq.~(\ref{eq:32}) (see Eq.~(\ref{eq:20}) for the case of $SU(2)$).

\section{Invariants due to local symmetries}

In this section we focus on the special case when the potentials $V_i(x)$ in the Super-Lagrangian are related to each other, for example,
through specific symmetries in finite domains of the embedding space.
To simplify the notation we work in one spatial dimension. However, our analysis can be easily extended to higher dimensions, as addressed
also at the end of section II. Moreover, 
according to the discussion in the previous section we focus without loss of generality exclusively on the case of the $SU(2)$ symmetry (two fields). We make the additional assumption that the two fields involved in the Super-Lagrangian, are stationary states corresponding to the same energy eigenvalue, i.e. although they obey different 
Schr\"{o}dinger problems of the same particle, they are eigenstates of the corresponding Hamiltonian operators with the same energy eigenvalue.  This condition is naturally met in quantum scattering with a continuous energy spectrum.

Firstly consider the stationary state
\begin{equation}
\mathbf{\Psi}(x,t)=e^{i E t} \mathbf{\Phi}(x)~~~~;~~~~\mathbf{\Phi}(x)=\begin{pmatrix}\Phi_1(x) \\ \Phi_2(x)  \end{pmatrix}
\label{eq:33}
\end{equation}
which exists due to the aforementioned assumption that the two Schr\"{o}dinger equations derived from the Super-Lagrangian (\ref{eq:3}) have at least one common energy eigenvalue. The most immediate scenario is to assume that the wave fields $\Psi_i(x,t)$ ($i=1, 2$) are scattering states at energy $E$ for the potentials $V_i(x)$ respectively. The continuity equation (\ref{eq:20}) for this particular state becomes
\begin{equation}
\frac{1}{2i} \frac{d}{dx}  \left( \Phi_2(x)^* \frac{d \Phi_1(x)}{d x} - \Phi_1(x) \frac{d \Phi_2^*(x)}{d x} \right) = i (V_2(x)-V_1(x))\Phi_1(x) \Phi_2^*(x) 
\label{eq:34}
\end{equation}
Let us now assume as a first case that in a domain $\mathcal{D}$ of coordinate space it holds: $V_2(x)=V_1(x)$ while for $x \notin \mathcal{D}$ we have $V_2(x) \neq V_1(x)$. Clearly, within this domain the two-field current
\begin{equation}
J_{12}=\frac{1}{2i}\left( \Phi_2(x)^* \frac{d \Phi_1(x)}{d x} - \Phi_1(x) \frac{d \Phi_2^*(x)}{d x} \right) = const., x \in \mathcal{D}
\label{eq:35}
\end{equation}
is independent of $x$ i.e. constant. Since the potential $V_2(x)$ within the domain $\mathcal{D}$ coincides with $V_1(x)$ the field $\Phi_2(x)$ will be a linear combination of two linearly independent solutions $\chi_1(x)$, $\chi_2(x)$ for the Schr\"{o}dinger problem corresponding to the potential $V_1(x)$. Without loss of generality we can assume that $\chi_1(x) \equiv \Phi_1(x)$. Thus, within the domain $\mathcal{D}$ we can write
\begin{equation}
\Phi_2(x)=c_1 \Phi_1(x) + c_2 \chi_2(x)
\label{eq:36}
\end{equation}
Inserting Eq.~(\ref{eq:36}) into the continuity equation (\ref{eq:34}) we find that the current
\begin{equation}
J_{\chi}=\frac{1}{2i} \left( \chi_2(x)^* \frac{d \Phi_1(x)}{d x} - \Phi_1(x) \frac{d \chi_2^*(x)}{d x} \right) = const., x \in \mathcal{D}
\label{eq:37}
\end{equation}
is constant within the domain $\mathcal{D}$. Due to the fact that $\Phi_1(x)$ and $\chi_2(x)$ obey the stationary Schr\"{o}dinger equation it is clear that $\Phi_1^*(x)$ and $\chi_2^*(x)$ obey it too. In this sense the constancy of $J_{\chi}$ within the domain $\mathcal{D}$ provides a generalization of the Wronskian valid for sub-domains of the 
embedding space. For scattering problems it holds that if $\Phi_1(x)$ is a (complex) solution, then, the other independent solution is $\Phi_1^*(x)$. As a consequence we obtain the standard conserved current $J=\frac{1}{2i} \left( \Phi(x)^* \frac{d \Phi(x)}{d x} - \Phi(x) \frac{d \Phi^*(x)}{d x} \right)$ involving $\Phi$ and $\Phi^*$ while the current involving only $\Phi$ is identically zero.

As a next step we discuss now the case of the presence of linear relations between the two potentials $V_1(x)$ and $V_2(x)$ in the 
sense that $V_2(x)=V_1(F(x))$ within the domain $\mathcal{D}$ where $F(x) = \sigma x + \rho$ with $\sigma = \pm 1; \rho \in \mathbb{R}$ is 
a linear transformation i.e. an inversion (parity) or translation.  Furthermore, we focus on the case when $V_1(F(x))=V_1(x)$ within $\mathcal{D}$.
This corresponds to the situation of the presence of a local symmetry in the domain $\mathcal{D}$. The theoretical description and
consequences of local symmetries as well as their impact on wave mechanical systems in acoustics, optics and quantum mechanics
have been discussed recently in a series of works \cite{Kalozoumis2014,LocSym}. Then, the continuity equation (\ref{eq:34}) within $\mathcal{D}$ becomes
\begin{equation}
\frac{1}{2i} \frac{d}{dx}  \left( \Phi_2(x)^* \frac{d \Phi_1(x)}{d x} - \Phi_1(x) \frac{d \Phi_2^*(x)}{d x} \right)=0
\label{eq:38}
\end{equation}
where $\Phi_2(x)$ is a solution of the Schr\"{o}dinger problem with the potential $V_1(F(x))$. Due to the symmetry $V_1(F(x))=V_1(x)$, valid within $\mathcal{D}$, the wave field $\Phi_2(x)$ can be $\Phi_1(F(x))$ or $\Phi^*_1(F(x))$ or any linear combination of them, provided that the transformation $F$ leaves the operator $\hat{H}_0$ in the Lagrangian (\ref{eq:3}) invariant. In the one-dimensional case which we focus on here, these are the inversion (parity) and translation transformations, while in higher dimensions also rotations, reflections and
permutation transformations are possible. Assuming that we restrict our analysis to the inversion and translation
transformations, we directly obtain the constancy, within the domain $\mathcal{D}$, of the currents
\begin{eqnarray}
\tilde{Q}&=&\frac{1}{2i} \frac{d}{dx}  \left( \Phi_1(F(x))^* \frac{d \Phi_1(x)}{d x} - \Phi_1(x) \frac{d \Phi_1^*(F(x))}{d x} \right) \nonumber \\
Q&=&\frac{1}{2i} \frac{d}{dx}  \left( \Phi_1(F(x)) \frac{d \Phi_1(x)}{d x} - \Phi_1(x) \frac{d \Phi_1(F(x))}{d x} \right)
\label{eq:39}
\end{eqnarray}
which are non-local, since they involve the field $\Phi_1(x)$ and its derivative at two different locations $x$ and $F(x)$, related by the transformation $F$. These are the currents derived in ref.\cite{Kalozoumis2014} using directly the Schr\"{o}dinger equation. In the present work we have shown that they are special cases resulting from the general continuity equation (\ref{eq:20}) after imposing specific constraints on the potentials $V_i(x)$ ($i=1, 2$). Here, we have developed a top-down approach for the derivation of these currents through a variational principle. As a by-product our treatment has led to the generalized continuity equation in Eq.~(\ref{eq:20}) connecting the wave fields of two different Schr\"{o}dinger problems.

We close this section with an example indicating some possible applications of the generalized continuity equation (\ref{eq:20}).
We again focus on the case of a single spatial dimension and furthermore, we assume that the potentials $V_1(x)$ and $V_2(x)$
possess the profile shown in Fig.~1.
\begin{figure}[h]
\centering
\includegraphics[width=0.95\textwidth]{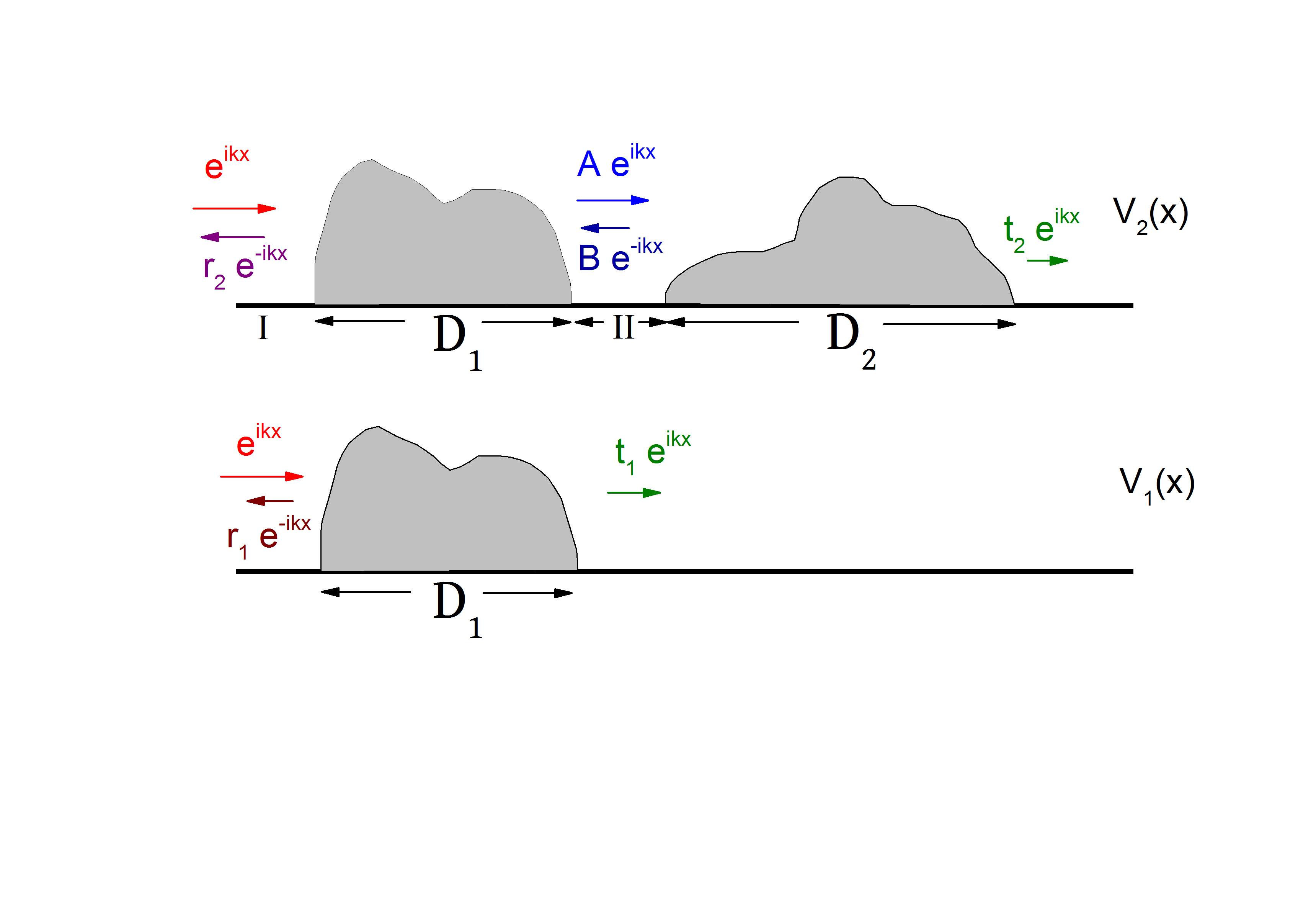}
\caption{The structure of the potential landscapes $V_1(x)$ and $V_2(x)$ used in our example. Within the domain $\mathcal{D}_1$ the two potentials are identical.  The potential in the domain $\mathcal{D}_2$ is arbitrary.}
\label{fig:fig1}
\end{figure} 

Within the domain $\mathcal{D}_1$ we have $V_1(x)=V_2(x)$. The potential $V_1(x)$ is zero outside the domain $\mathcal{D}_1$ while $V_2(x)$ is zero outside $\mathcal{D}_1 \cup \mathcal{D}_2$. The shape of $V_2(x)$  within the domain $\mathcal{D}_2$ is arbitrary. We will show in the following that it is possible to determine the intermediate coefficients $A$ and $\vert B \vert$ in terms of the reflection and transmission
coefficients $r_1$, $t_1$ and $r_2$ (see Fig.\ref{fig:fig1}).
The current $J_{12}$ (see Eq.~(\ref{eq:35})) is constant in the complete region, starting from the left edge of domain $\mathcal{D}_1$ 
up to the left edge of domain $\mathcal{D}_2$.
It is straightforward to derive its value separately in region I and region II 
\begin{equation}
J_{12}^{(I)}=k (1 - r_1^* r_2) = J_{12}^{(II)}=k t_1^* A 
\label{eq:40}
\end{equation}
This relation determines the coefficient $A$ as
\begin{equation}
A=\frac{1- r_1^* r_2}{t_1^*}
\label{eq:41}
\end{equation}
Employing the usual current conservation in the potential $V_2(x)$ we determine also the value of $\vert B \vert$ as
\begin{equation}
\vert B \vert = \sqrt{ \vert \frac{1-r_1^* r_2}{t_1^*} \vert^2 -(1 - \vert r_2 \vert^2)}
\label{eq:42}
\end{equation}
Thus, up to the phase of $B$, the solution in the interior of the landscape $V_2(x)$ (region II) is determined by the output obtained by the asymptotic scattering properties of these two different Schr\"{o}dinger problems. We also see that to achieve transparency through the domain $\mathcal{D}_1$ for the potential $V_2(x)$ we have to obey the condition
\begin{equation}
\vert r_2 \vert = \sqrt{1 - \vert \frac{1-r_1^* r_2}{t_1^*} \vert^2}
\label{eq:43}
\end{equation}
which is implicit with respect to $r_2$.

\section{Summary and conclusions}

We have established a variational approach based on a Super-Lagrangian which leads to generalized continuity equations linking pairwise the wave fields belonging to different Schr\"{o}dinger problems. Our focus has been on the case of different external potentials. A variety of properties can be derived based on these generalized continuity equations. We have addressed the emergence of conservation laws (spatially invariant currents) occurring in finite subdomains of the embedding space whenever the two potentials in the pair coincide. In a first simple application the usefulness of the derived continuity equations has been demonstrated. The investigation of more complex setups is envisaged in a future work. In particular, it would be interesting to generate dualities between the solutions of different Schr\"{o}dinger problems for interacting many-body systems.

{}

\end{document}